\documentclass[prl,nofootinbib,twocolumn,floatfix,showpacs]{revtex4}
\usepackage{graphicx}
\usepackage{amsmath}
\usepackage{amssymb}
\bibpunct[, ]{[}{]}{,}{a}{,}{,}
\ifnum\lefthyphenmin<2\lefthyphenmin=2\fi
\ifnum\righthyphenmin<2\righthyphenmin=2\fi

\begin{document}
\title{A scale-free network hidden in the collapsing polymer}

\author{A. Kabak\c c\i o\u glu}\email{alkan@pd.infn.it}
\author{A. L. Stella}\email{stella@pd.infn.it}
\affiliation{INFM-Dipartimento di Fisica and Sezione INFN, Universit\`a di Padova, I-35131 Padova, Italy}

\date{\today }

\begin{abstract}
We show that  the collapsed globular phase of  a polymer accommodates a
scale-free   incompatibility  graph  of its contacts. The  degree
distribution of this network is found to decay with the exponent
$\gamma = 1/(2-c)$ up to a cut-off degree $d_c \propto L^{2-c}$,
where $c$  is the loop exponent for dense polymers
($  c  =  11/8$  in  two  dimensions) and $L$ is the length of the polymer.
Our results exemplify how  a scale-free network (SFN) can emerge from standard
criticality.
\end{abstract}

\pacs{89.75.-k, 64.60.Fr, 36.20.Ey}

\maketitle

The underlying network
structures of the world-wide-web,
power grids, social contacts, protein interactions, etc.\cite{internet}
bear peculiar properties absent in random networks.
The graphs associated with them not only share a power law degree distribution
commonly referred as ``scale-free'', but also
an unusually small diameter (a measure of the average shortest distance between
two nodes) \cite{small-world} and typically a high degree of clustering.
The advantages/disadvantages of these qualities
on the efficiency of the network have been thoroughly discussed in the context of
epidemic spreading \cite{epidemic} and resilience against
random or malicious attacks \cite{survival}.

The search for non-growing SFNs originating from
equilibrium statistics and/or optimization has been focus of attention
recently \cite{static_opt,static_landscape,static_marco}.
And yet, although some statistical models on SFNs have already been studied
\cite{sandpile,ising,rw,saw}, there is, to our knowledge, no demonstrated link
from conventional critical models to the criticality hinted by the network's
scale-freeness.
As shown here, this link may be established by associating a network with each
microstate of the model.  Such a network representation is also convenient for studying
the system's complexity and may have wider applicability. 
For example, a recipe similar to ours was recently adopted
for investigating the navigational complexity of cities \cite{sneppen}.

In this Letter, we demonstrate that a SFN is associated with the 
collapsed phase at low temperature, $T$, and presumably also with
the $\theta$-point ($T=T_\theta$) of a polymer.
By modelling the collapsed polymer as a self-attracting self-avoiding walk \cite{Carlo}
on a square lattice we obtain the degree exponent in terms of a critical index known exactly
for 2D. Numerical simulations confirm the analytically predicted degree exponent and show
that the scale-free character of the network is a
direct consequence of the polymer criticality.

\begin{figure}[t]
\begin{center}
\end{center}
\includegraphics[width=8cm]{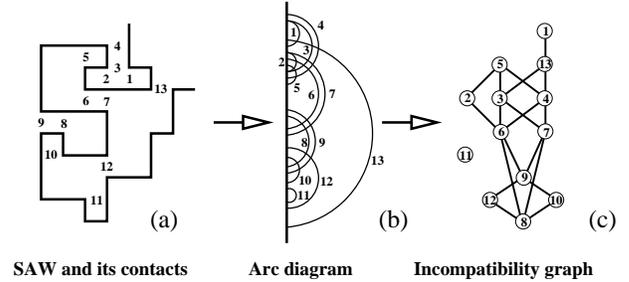}
\caption[]{A polymer on a square lattice and its self-contacts
together with the corresponding arc-diagram and the incompatibility graph.}
\label{arc-diagram}
\end{figure}
The configuration of a polymer chain of $L$ monomers can be represented
by an arc-diagram (see Fig.~\ref{arc-diagram}a,b) which carries the same information
as the contact map \cite{vendruscolo}.
It is obtained when one imagines
stretching the polymer into a horizontal straight line
while pairs of monomers that were originally in contact
are joined by arcs on the upper half plane (the diagram is drawn on a
plane even though the polymer may be embedded in arbitrary dimensions).
The arc-diagram is said to be {\it planar} if no two arcs cross each other, i.e.,
if there are no interpenetrating contacts.
Otherwise, a non-trivial incompatibility graph (IG) can be associated with
the arc-diagram (Fig.~\ref{arc-diagram}c), such that, each arc (contact) represents a node on the
IG and two nodes are connected by an edge if the corresponding arcs cross.
More precisely, if the monomer units of the polymer are
indexed $1$..$L$ and the contacts are labeled by indices of the involved
monomer pair $(i,j)$ with
$i<j$, then the two nodes $(i_1,j_1)$ and $(i_2,j_2)$ of the IG are joined by
an edge if and only if there exists an integer $n$ satisfying $i_1<n<j_1$ and $i_2<n<j_2$.

The IG provides a representation for the complexity of the contact structure of a
folded polymer with particular reference to the arrangement of pseudoknots.
This representation is frequently
used in RNA folding research \cite{haslinger,baiesi,kabakcioglu} where
pseudoknots are biologically active structural components.
An empty IG (no edges, meaning no pseudoknots) is a common simplifying
assumption in RNA prediction algorithms \cite{Nussinov,Zuker,bundschuh-hwa}.
Alternatively, in case of the traveling salesman problem (TSP) the
same complexity refers to the
shortest path connecting a given number of cities on a map where repeated
visitations of a neighborhood compounds to a contact. In d=2, TSP has recently been
conjectured to be in the same universality class as dense polymers \cite{tsp}.
Rosvall {\it et al.} use a similar
mapping from a city road-map to the IG, through which they quantify the
information content \cite{sneppen}.

Our goal here is to show that the IG of a homopolymer at
low temperatures is a scale-free network. To this end,
we will approximate the true arc-diagram of the polymer with
a random arc-diagram where the probability
of having an arc between two monomers $m$ distance apart along the
chain is given by its asymptotic value for an infinite polymer, i.e.,
$P(m) \sim m^{-c}$. The exponent $c$ is related to the return probability
and equals $d/2$ for a random walk \cite{Redner}. Its counterpart for the self-avoiding
polymer is $c = d\nu - \sigma_4$ \cite{Duplantier,DupPRL}. Note that, apart from the modified radius of gyration
exponent $\nu$ ($\nu = 1/2$ for the random walk) it additionally involves the four-leg vertex
exponent $\sigma_4$ which, e.g., plays a key role in DNA physics
\cite{Mukamel}. The exponents $\nu$ and $\sigma_4$ are known exactly or to high precision
for all integer dimensions, except for $T<T_\theta$ in 3D \cite{Carlo}.
We will find below that the tail of the degree distribution for the IG of such a
random arc-diagram obeys the power law
\begin{equation}
\label{deg_scaling}
P(d) \sim d^{-\gamma}\ \ ,
\end{equation}
with $\gamma = 1/(2-c)$ in an infinitely long polymer.

Consider an arc-diagram of the polymer with the discrete monomer
centers labeled by integers $1..L$. As discussed above, we set the
arc-length distribution to be
\begin{equation}
\label{pdf}
P(m) =  m^{-c} / (\zeta(c,1)-\zeta(c,L+1)) \ ,
\end{equation}
where the proper normalization is given in terms of
the generalized (Hurwitz) zeta function
\begin{equation}
\label{zeta-def}
\zeta(c,a) \equiv \sum_{n=0}^{\infty}{(n+a)^{-c}} \ .
\end{equation}
Throughout the paper we will
take $c>1$, as is the case for all systems of interest.
We start by considering a single arc in the arc-diagram, corresponding
to a loop of length $m$. We label the
monomers lying on the loop as $1 < i < m-1$, so that the contacting monomer
pair $(0,m)$ which closes the loop is excluded. We want
to calculate the probability distribution for the corresponding node in the
equivalent incompatibility graph to have degree $d$.

The probability that the considered arc crosses another one which has one
of its legs at monomer $i$ ($1 < i < m-1$) is
\begin{equation}
\label{pi}
\pi_{i,m} = \frac{1}{2}\frac{\zeta(c,i+1) + \zeta(c,m-i+1)}{\zeta(c,1)-\zeta(c,L+1)}\ , 
\end{equation}
Below, we fix the average density of contacts
(which is a function of temperature only) through the arc fugacity, $z$,
so that the arc density satisfies
\begin{equation}
\label{fugacity}
\sigma = z\, \big [ \partial{(1-z)^{-1}}/\partial{z} \, \big ]  / (1-z)^{-1} = z/(1-z)\ .
\end{equation}

The probability that out of the $n$ arcs originating from the site $i$,
$d_i$ will cross the reference arc ($0,m$) is
\begin{eqnarray}
\label{Pi}
\Pi_{i,m}(d_i) = & \nonumber\\
 (1-z) \, \sum_{n=d_i}^{\infty} & { z^n \,
                   \binom{n}{d_i}\, \pi_{i,m}(c)^{d_i}\, 
                   [1-\pi_{i,m}(c)]^{n-d_i}}\ .
\end{eqnarray}
Now we can write down the probability that the reference arc crosses a total of $d$ arcs as:
\begin{equation}
\label{Pd}
P_m(d) = \sum_{\{d_i\}}^{\sum_i{d_i}=d}{ \prod_{i=1}^{m-1}{\Pi_{i,m}(d_i)}} \ .
\end{equation}
The difficulty due to the restricted sum is overcome by shifting to the
grand-canonical formulation with an arc-crossing fugacity, $\mu$. If we
define $\tilde{P}_m(\mu)$ to be the Laplace transform of Eq.(\ref{Pd}), then
a further summation over all loop lengths $m$ gives the Laplace transform
$\tilde{P}(\mu,L)$ of the degree distribution we are after. Using Eqs.(\ref{Pi},\ref{Pd})
we obtain
\begin{eqnarray}
\label{goal}
\tilde{P}(\mu,L) &=& \sum_{d=0}^{\infty}{P(d,L)\,\mu^d} \nonumber\\
&=& \sum_{m=1}^{L}{m^{-c}\,\prod_{i=1}^{m-1}{[ 1 + \sigma (1-\mu) \pi_{i,m} ]^{-1}}}
\end{eqnarray}

\begin{figure}[t]
\begin{center}
\end{center}
\includegraphics[width=8cm]{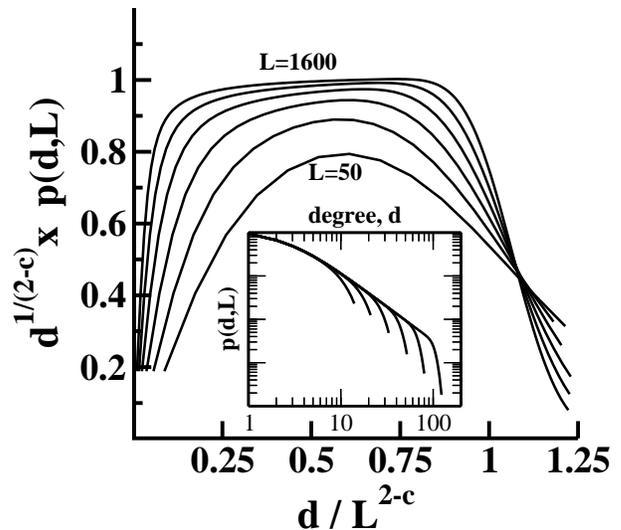}
\caption[]{
The plots for increasing $L$ converge to the expected scaling function of Eq.(\ref{fx}).
The inset shows the degree distribution for different chain lengths before scaling the axes.}
\label{fig-collapse}
\end{figure}
We solved for $P(d,L)$ numerically using Mathematica. The probability distribution for
the node degrees obtained exhibits a clear power law with a sharp finite size cut-off
(Fig.~\ref{fig-collapse} inset). The analysis of the data for a range of lengths $L$
and several values of $c$ gives the finite size scaling form:

\begin{equation}
\label{scaling-equation}
P(d,L) = d^{-1/(2-c)}\,f(d/L^{2-c})
\end{equation}
where $f(x)$ is a scaling function determining the window of validity for
the power law.
\begin{equation}
\label{fx}
f(x) =
\begin{cases}
const, & x \lesssim O(1)\\
0, & \mbox{otherwise}.
\end{cases}
\end{equation}
When the two axes are normalized in accordance with
Eq.(\ref{scaling-equation}), we obtain a convincing asymptotic collapse for $P(d,L)$ for
increasing $L$ (Fig.~\ref{fig-collapse}), hence assuring the validity of the
assumed scaling form.

It is clear from Eqs.(\ref{scaling-equation},\ref{fx}) that the power law
disappears when $c \ge 2$. This, in fact, is the case for a freely fluctuating
self-avoiding coil in two and three dimensions. For convenience, we list below
the value of the loop exponent, $c$, for various cases:

\begin{center}
\begin{tabular}{c|c|cl}
\label{tab1}
Temperature & \ 2D & \ 3D &\\\hline
$T < T_\theta$ & \ 1.375 \  & \ - &\\
$T = T_\theta$ & \ 1.571 \  & \ 1.5&\\ 
$T > T_\theta$ & \ 2.688 \  & \ 2.22&\mbox{\cite{pade-borel}} \\
\end{tabular}
\end{center}
For the low T phase in three dimensions no estimate of $\sigma_4$ is
available yet. In this case, simulations suffer from severe surface effects which
add to the difficulties of sampling compact configurations.

\begin{figure}[t]
\includegraphics[width=7cm]{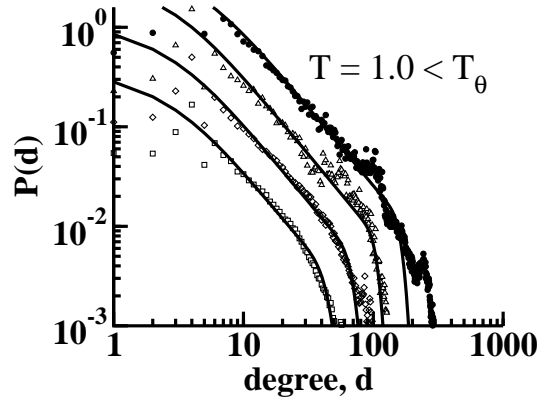}
\caption[]{Averaged degree distribution for the IG of
a self-attracting self-avoiding walk of 400 ($\Box$),800($\Diamond$),1600($\triangle$),
3200($\bullet$) steps on square lattice at $T/\epsilon=1.0 < T_\theta/\epsilon$.
The solid curves are the model's predictions with $c = d\nu -\sigma_4 = 1.375$.}
\label{fig-degree-dist}
\end{figure}

The random-graph approximation we employed above omits correlations among contacts in the IG graph.
For example, we ignored the fact that the IG is bipartite in two dimensions  \cite{haslinger}.
Thus, the conclusions  of the random-graph scenario are not guaranteed a priori to hold
for true  polymers and it is necessary to confirm them through numerical simulations.
For this purpose, we generated
self-avoiding walks on a square lattice up to length $L=3200$ by using the
Pruned-Enriched Rosenbluth Method (PERM) \cite{perm}, a particularly
efficient Monte-Carlo technique for interacting polymers.
A nearest neighbor attractive interaction $\epsilon$ was introduced in order to
induce collapse \cite{Carlo}. For each sample we obtained the normalized degree distribution of its
incompatibility graph and averaged them with their associated
Boltzmann weight (Fig.~\ref{fig-degree-dist}).

According to the table above and Eq.(\ref{scaling-equation}), high $T$ polymers do not have
scale-free IGs. We first checked through computer simulations
at $T > T_\theta$ that this indeed is true. In fact, the IG in this case is composed
of many disconnected subgraphs with
an exponentially decaying distribution of number of nodes.
Our preliminary results show that the $c$ exponent controls the power-law distribution
for the span along the polymer backbone of each isolated subgraph comprising the IG.

According to Eqs.(\ref{scaling-equation},\ref{fx}) SFN should emerge when $c$ gets smaller
than 2, i.e. when we cross the $\theta$-point.
We confirm this by considering next the case where the scale-free nature of the IG
is expected to be most easy to demonstrate, namely the $T<T_\theta$ regime in two dimensions.
The small loop exponent $c$ in this regime renders a wider scaling region on the basis of
Eq.(\ref{scaling-equation}).
The temperature was fixed to
$T/\epsilon = 1.0 < T_\theta/\epsilon \simeq 1.54 $ \cite{Carlo}, deep in the low T phase.

The degree distribution for not too high IG node degrees appears to be
self-averaging, since we were able to identify the power law behavior even in
a single sample. The data presented in Fig.~\ref{fig-degree-dist} are averaged
over at least 1000 independent samples, mainly to determine
the cut-off behavior. For comparison, the
corresponding random-graph prediction obtained from Eq.(\ref{goal})
is also shown in the same figure with a solid curve. The curve and the data points
are overlapped
by a suitable {\it vertical} shift to compensate for the discrepancy between
the two at very small degree numbers ($d<5$): The window $0\le d < 5$ is
irrelevant for scaling purposes, yet it accounts for a finite fraction of
the probability integral. An incorrect prediction in this region (due to
deviations from the asymptotic form in Eq.(\ref{pdf})) consequently
results in a finite vertical shift on the right-hand side of the distribution.

Fig.~\ref{fig-degree-dist} shows the degree distributions obtained from simulations
of chains of various lengths.
The agreement with the random-graph estimate calculated above
is also shown (solid line). Not only the degree exponent, $\gamma = 1/(2-c) = 1.6$, which is
given by the slope in the log-log plot, but also the $L$ dependence of the finite size cut-off
is well predicted by Eqs.(\ref{scaling-equation},\ref{fx}).

Although the $\theta$-point, too,
appears as a good candidate with a predicted network exponent of $\gamma = 2.33 (2.0)$
for 2D(3D), relatively longer polymers required in order to obtain a convincing power law
scaling make simulations harder due to either self-trapping or strong logarithmic corrections,
respectively.

Most known examples of SFNs also exhibit a high degree of clustering. Our graphs being ideally
bipartite, the standard clustering coefficient based on counting triangles vanishes
(except for a small contribution from buried polymer ends).
Instead, we considered the probability that two nodes have a common nearest neighbor
given that they already share another one and checked how it differs from its expected
value for two uncorrelated nodes with the same degree. The latter probability
is given by
\begin{equation}
\label{clustering}
P_{rand}(d_1,d_2) = 1 - \frac{(N-d_1-1)...(N-d_1-d_2)}{(N-1)(N-2)...(N-d_2)}\ ,
\end{equation}
where N is the number of nodes in the subgraph excluding the two nodes with degrees $d_1,d_2$.
Note that the Eq.(\ref{clustering}) is symmetric under the exchange of degrees.
We found that $\langle P(d_1,d_2) - P(d_1,d_2)_{rand} \rangle \simeq 0.16$,
indicating a high degree of clustering. A feature possibly due to the spatial proximity of
the loops corresponding to the two nodes that share a common nearest neighbor.

In conclusion, we showed that a collapsed polymer globule in two dimensions
accommodates a scale-free network.
By means of a model based on the critical properties
of the polymer, we conjecture exact exponents for the scaling of the degree distribution.
The model predicts similar behavior also at the $\theta$-point, but with different degree
exponents. Resulting networks are scale-free upto a degree number obtained by finite
size scaling.  The cut-off is determined by a power of $L$ which is the inverse
of the degree exponent $\gamma$, such that the power law scaling is less
prominent (cut-off is more severe) at higher values of $\gamma$.
It is conceivable that a similar mechanism applying to other known examples of SFNs
may serve as a sieve selecting low degree exponents.
We note that Barab\'{a}si-Albert networks \cite{internet} exhibit similar finite size
effects \cite{amaral}.

The connection established here between polymers at low temperature
and the scale-free networks will prove fruitful in other related problems
as well. One possible connection is to the TSP \cite{tsp}, where
the topology of the related IG could serve as a checkpoint for the conjectured
correspondence with dense polymers.
One is also tempted to attribute to the IG structure of a folded biopolymer,
such as a protein in its native state, a signature of the folding
kinetics. For example, since the IG gives a measure of how ``inter-penetrating''
the contacts are, a correlation between high-degree nodes and the earlier
formed contacts or the folding nucleus \cite{nucleus} is plausible.

This work was supported by Italian MIUR through FIRB-2001 and PRIN-2003.

\end{document}